\documentclass[prd,amsmath,amssymb,superscriptaddress,nofootinbib,showpacs]{revtex4}
\usepackage{graphicx}
\usepackage{dcolumn}
\usepackage{amsmath}
\usepackage{epsfig}
\usepackage{times}
\usepackage{amsmath}
\usepackage{amsfonts}
\usepackage{amssymb}
\usepackage{amsthm}
\usepackage{bm}
\usepackage{latexsym}
\usepackage{graphicx}
\usepackage{color}
\usepackage{url}
\usepackage{subfigure}
\usepackage{ulem}
\long\def\inst#1{\par\nobreak\kern 4pt\nobreak
    {\itshape #1}\par\vskip 10pt plus 3pt minus 3pt}
\RequirePackage{xspace}

\def\babar{\mbox{\slshape B\kern-0.1em{\smaller A}\kern-0.1em
    B\kern-0.1em{\smaller A\kern-0.2em R}}}
\def\Kbar    {\kern 0.18em\overline{\kern -0.18em K}{}\xspace}

\def\Kz      {\ensuremath{K^0}\xspace}
\def\Kzb     {\ensuremath{\Kbar^0}\xspace}
\def\KzKzb   {\ensuremath{\Kz {\kern -0.16em \Kzb}}\xspace}

\def\Ks     {\ensuremath{K_S}\xspace}
\def\Kl     {\ensuremath{K_L}\xspace}
\def\KsKs   {\ensuremath{\Ks {\kern -0.16em \Ks}}\xspace}
\def\KlKl   {\ensuremath{\Kl {\kern -0.16em \Kl}}\xspace}
\def\KsKl   {\ensuremath{\Ks {\kern -0.16em \Kl}}\xspace}
\def\KlKs   {\ensuremath{\Kl {\kern -0.16em \Ks}}\xspace}
\def\Dbar    {\kern 0.18em\overline{\kern -0.18em D}{}\xspace}
\def\Dz      {\ensuremath{D^0}\xspace}
\def\Dzb     {\ensuremath{\Dbar^0}\xspace}
\def\DzDzb   {\ensuremath{\Dz {\kern -0.16em \Dzb}}\xspace}

\newcommand{\DsP}{\ensuremath{D_s^+}\xspace}
\newcommand{\DsM}{\ensuremath{D_s^-}\xspace}
\newcommand{\DspDsm}{\ensuremath{\DsP {\kern -0.16em \DsM}}\xspace}
\newcommand{\Dp}{\ensuremath{D^+}\xspace}
\newcommand{\Dm}{\ensuremath{D^-}\xspace}
\newcommand{\DpDm}{\ensuremath{\Dp {\kern -0.16em \Dm}}\xspace}

\def\Bbar    {\kern 0.18em\overline{\kern -0.18em B}{}\xspace}

\def\Bz      {\ensuremath{B^0}\xspace}
\def\Bzb     {\ensuremath{\Bbar^0}\xspace}
\def\BzBzb   {\ensuremath{\Bz {\kern -0.16em \Bzb}}\xspace}
\def\Bu      {\ensuremath{B^+}\xspace}
\def\Bub     {\ensuremath{B^-}\xspace}

\def\BpBm    {\ensuremath{\Bu {\kern -0.16em \Bub}}\xspace}

\def\Dp      {\ensuremath{D^+}\xspace}

\newcommand{\optbar}[1]{\shortstack{{\tiny (\rule[.4ex]{1em}{.1mm})}
  \\ [-.7ex] $#1$}}
\def\BorBbar    {\kern 0.18em\optbar{\kern -0.18em B}{}\xspace}
\def\DorDbar    {\kern 0.18em\optbar{\kern -0.18em D}{}\xspace}
\def\KorKbar    {\kern 0.18em\optbar{\kern -0.18em K}{}\xspace}

\def\pep2{PEP-II}
\mathchardef\Upsilon="7107
\def\Y#1S{\ensuremath{\Upsilon{(#1S)}}\xspace}

\begin{document}

\title{\large \bfseries \boldmath Prospects for rare and forbidden hyperon decays at BESIII }
\author{Hai-Bo Li}\email{lihb@ihep.ac.cn}
\affiliation{Institute of High Energy Physics, Beijing 100049, People's Republic of China}
\affiliation{University of Chinese Academy of Sciences, Beijing 100049, People's Republic of China}


\date{\today}

\vspace{0.4cm}
\begin{abstract}
    The study of hyperon decays at the Beijing Electron Spectrometer III (BESIII) is proposed to investigate the events of $J/\psi$ decay into hyperon pairs, which provide a pristine experimental environment at the Beijing Electron--Positron Collider II.  About $10^{6}$--$10^8$ hyperons, i.e., $\Lambda$, $\Sigma$, $\Xi$, and $\Omega$, will be produced in the $J/\psi$ and $\psi(2S)$ decays with the proposed data samples at BESIII.  Based on these samples, the  measurement sensitivity of the branching fractions of the hyperon decays is in the range of $10^{-5}$--$10^{-8}$.  In addition, with the known center-of-mass energy and ``tag technique,''  rare decays and decays with invisible final states can be probed. 
\end{abstract}


 \maketitle

\section{Introduction}

 In this paper, the hyperon physics program at the Beijing Electron Spectrometer III/Beijing Electron--Positron Collider II (BESIII/BEPCII) and prospects of its upgradation were surveyed.
  Even with the designed luminosity of the BEPCII,  $\sim$$10^{10}$ $J/\psi$ and $\psi(2S)$ events can be collected with a running period of one year.
  The inclusive hyperon production rate can reach a few percent in the $J/\psi$ decays, while the hyperon-pair production rate is $\sim$$10^{-3}$.
   With these data,  the decay properties of the spin-1/2 baryon octet can be revisited and the decay parameters with the coherent production
   of hyperon pairs in $J/\psi \to B \bar{B}$ decays (where the spin-1/2 hyperon pairs are in the $1^{--}$ state) can be probed. This survey will focus on hyperon decays that can be studied
   at BESIII and rare decays, for which  the search has not yet begun.

\section{Status of BESIII }

The BESIII detector consists mainly of a cylindrical main draft
chamber (MDC) with momentum resolution $\sigma_{p_t}/p_t \sim
0.5\%$ for charged particles with momentum at 1.0 GeV, a
time-of-flight (TOF) system with two layers of plastic
scintillator counters located outside of the MDC, and a highly hermetic
electromagnetic calorimeter (EMC) with an energy resolution of
$\sigma_E /E = 2.5\%/\sqrt{E(\mbox{GeV})}$~\cite{Ablikim:2009aa}.
The MDC has its first sensitive layer at a radius of 6.0 cm from
the interaction point (IP), and the MDC combined with a magnetic field of
1.0 T provides precise momentum measurements of charged
particles with transverse momentum $>$50 MeV.

BESIII has so far acquired $\sim$$5\times 10^{8}$ and
$\sim$$13\times 10^{8}$ events on the $\psi(2S)$ and $J/\psi$ peaks, respectively.
In the next few years, by the end of 2018, $\sim$$10^{10}$ events on the  $J/\psi$
peak are going to be collected, and subsequently $\sim$$3\times 10^{9}$ events on the $\psi(2S)$ peak will be taken.
 The expected data samples collected at BESIII/BEPCII per year
are summarized in Table~\ref{tab:1}~\cite{Asner:2008nq}. In this paper, the
sensitivity studies are based on 5 fb$^{-1}$ luminosity at the
$J/\psi$ or $\psi(2S)$ peaks for probing hyperon decays.
\begin{table}[htbp]
\caption{$\tau$-Charm production at BEPCII in one-year run
($10^7$ s).} \label{tab:1}
\begin{tabular}{@{}lll}
\hline
               & Central-of-Mass    & \#Events  \\
Data Sample    & (MeV)                   & per year  \\
\hline
$J/\psi$ &  3097     & $10\times 10^9$\\
$\tau^+\tau^-$   & 3670  & $12\times 10^6$ \\
$\psi(2S)$ & 3686  & $3.0\times 10^9$ \\
$\DzDzb$ & 3770  & $18\times 10^6$ \\
$\DpDm$ & 3770  & $14\times 10^6$ \\
$\DspDsm$ & 4030  & $1.0\times 10^6$ \\
$\DspDsm$ & 4170  & $2.0\times 10^6$ \\
\hline
\end{tabular}
\end{table}

\begin{table}[htbp]
\caption{Hyperon production from the  $J/\psi$ or  $\psi(2S)$ two-body decays with  $10^{10}$ events on the  $J/\psi$
peak and  $3\times 10^{9}$ events on the $\psi(2S)$ peak. $N_B$ is the number of the expected hyperon events. Data are from the Particle Data Group (PDG2016)~\cite{pdg2016}.} \label{tab:2}
\begin{tabular}{@{}lll}
\hline
Decay mode    &  ${\cal B}(\times 10^{-3})$                  & $N_B$ ($\times 10^6$) \\
\hline
$J/\psi\to \Lambda \bar{\Lambda}$ & $1.61\pm 0.15$      & $16.1\pm 1.5$ \\
$J/\psi\to \Sigma^0\bar{\Sigma}^0$ &$1.29\pm 0.09$ &  $12.9\pm 0.9$ \\
$J/\psi\to \Sigma^+\bar{\Sigma}^-$ &$1.50\pm 0.24$ & $15.0\pm 2.4$  \\
$J/\psi\to \Sigma(1385)^-\bar{\Sigma}^+$ (or c.c.) &$0.31\pm 0.05$ & $3.1\pm 0.5$ \\
$J/\psi\to \Sigma(1385)^-\bar{\Sigma}(1385)^+$ (or c.c.) &$1.10\pm 0.12$ & $11.0\pm 1.2$  \\
$J/\psi\to \Xi^0\bar{\Xi}^0$ &$1.20\pm 0.24$ & $12.0\pm 2.4$  \\
$J/\psi\to \Xi^-\bar{\Xi}^+$ &$0.86\pm 0.11$ & $8.6\pm 1.0$ \\
$J/\psi\to \Xi(1530)^0\bar{\Xi}^0$ &$0.32\pm 0.14$ & $3.2\pm 1.4$ \\
$J/\psi\to \Xi(1530)^-\bar{\Xi}^+$ &$0.59\pm 0.15$ & $5.9\pm 1.5$ \\
$\psi(2S)\to \Omega^-\bar{\Omega}^+$ &$0.05\pm 0.01$  &$0.15 \pm 0.03$  \\
\hline
\end{tabular}
\end{table}

\begin{table}[htbp]
\caption{Hyperon production from the  $J/\psi$  three-body  decays with  $10^{10}$ events on the  $J/\psi$
peak and  $3\times 10^{9}$ events on the $\psi(2S)$ peak. $N_B$ is the number of the expected hyperon events. Data are from PDG2016~\cite{pdg2016}.} \label{tab:3}
\begin{tabular}{@{}lll}
\hline
Decay mode    &  ${\cal B}(\times 10^{-4})$                  & $N_B$ ($\times 10^6$) \\
\hline
$J/\psi\to pK^- \bar{\Lambda}$ & $8.9\pm 1.6$      & $8.9\pm 1.6$  \\
$J/\psi\to \Lambda \bar{\Lambda} \pi^+\pi^-$ & $43\pm 10$      & $43\pm 10$  \\
$J/\psi\to pK^- \bar{\Sigma}^0$ & $2.9\pm 0.8$      &  $2.9\pm 0.8$ \\
$J/\psi\to \Lambda \bar{\Sigma}^- \pi^+$ (or c.c.) & $8.3\pm 0.7$      & $8.3\pm 0.7$  \\
$J/\psi\to \Lambda \bar{\Sigma}^+ \pi^-$$^*$ (or c.c.) & $8.3\pm 0.7$      & $8.3\pm 0.7$ \\
$J/\psi\to pK^- \bar{\Sigma}(1385)^0$ & $5.1\pm 3.2$      & $5.1\pm 3.2$ \\
\hline
\end{tabular}
\vspace{-0.2cm}
  \begin{flushleft}
  \footnotesize{$^{*}$  Estimated from isospin symmetry. }
   \end{flushleft}
\end{table}

As indicated in Table~\ref{tab:2}, the branching fractions for the $J/\psi$ decays into hyperon pairs are on the order of $10^{-3}$, and $\sim$$10^7$ hyperon pairs can be produced in the $J/\psi$ decays per year. The $\Omega^-$ can be only produced in the $\psi(2S)$ decays owing to the allowed phase space, and $\sim$$10^4$--$10^5$ $\Omega^- \bar{\Omega}^+$ pairs will be produced per year at BESIII.  Since there is always a neutron or a neutrino produced from the $\Sigma^-$ hadronic decays or semileptonic decays,  one has to use a ``tag technique'' to look for the $\Sigma^-$ decays. In Table~\ref{tab:3},  the three-body decays of $J/\psi\to \bar{\Lambda} \Sigma^- \pi^+ + {\rm c.c.}$ can be used to study the $\Sigma^-$ decays by looking at the recoiling mass of the $\bar{\Lambda} \pi^+$.


 \section{Semileptonic  hyperon decays}
\label{sec:semi}

The Cabibbo--Kobayashi--Maskawa matrix elements $|V_{ud}|$ and $|V_{us}|$ characterize quark mixings in 
$d \to u e^- \bar{\nu}_e$ and $s \to u e^- \bar{\nu}_e$ processes~\cite{Cabibbo:1963yz,Kobayashi:1973fv} in the Standard Model (SM).
So far the most precise determinations of  $|V_{ud}|$ and $|V_{us}|$ have been obtained, respectively, from super-allowed Fermi transitions
together with pion decays and from leptonic and semileptonic
kaon decays~\cite{meson-v,meson-v1,meson-v2}. However, hyperon semileptonic
decays  can also provide independent constraints
on $|V_{ud}|$ and $|V_{us}|$~\cite{hyperon-v,hyperon-v1}.

  In addition one can test the $V$--$A$ structure~\cite{Weinberg:2009zz} of the charged currents in the semileptonic hyperon decays~\cite{hyperon-v1,Chang:2014iba}.  The semileptonic hyperon decays provide essential
information on the structures of the nucleon and low-lying
hyperons. The data for the semileptonic hyperon decays reveal experimentally the pattern
of flavor SU(3) symmetry breaking~\cite{Pham:2012db}. In exact flavor SU(3)
symmetry, the ratios of the axial-vector and vector constants
$g_1/f_1$ are expressed only by the two constants $F$ and $D$.
Similarly, those of the vector constants $f_2/f_1$ are written in
terms of the anomalous magnetic moments of the proton and
the neutron with flavor SU(3) symmetry assumed~\cite{hyperon-v}.  Details of the semileptonic hyperon decay constants can be found in a review paper~\cite{hyperon-v}, and recent developments are described elsewhere~\cite{Yang:2015era,Faessler:2008ix,Borasoy:1998pe,Geng:2009ik,Ledwig:2014rfa}.  However,
the experimental data for the semileptonic hyperon decays show that the flavor SU(3)
symmetry may be manifestly broken~\cite{Yang:2015era}, and further precision data are needed, especially for the decays of  $\Xi^0$,  $\Xi^-$, and $\Omega^-$ as listed in Table~\ref{tab:beta-decays}.  The current measured values for  the form-factor ratio $g_1(0)/f_1(0)$ in the Cabibbo model~\cite{Bourquin:1985su}
are also listed in Table~\ref{tab:beta-decays}.

     The decays of $\Sigma^- \to \Sigma^0 e^- \bar{\nu}_e$ and $\Xi^- \to \Xi^0  e^- \bar{\nu}_e$ have not been observed yet.
 As the lepton pairs ($e^- \bar{\nu}_e$) are too soft to be detected by the BESIII detector, to study  $\Sigma^- \to \Sigma^0 e^- \bar{\nu}_e$  in the $J/\psi\to \bar{\Lambda} \Sigma^- \pi^+ $ decay $, \bar{\Lambda}  \pi^+$ can be fully reconstructed as the ``tag  side'',  following an examination of the recoiling mass of the  $ \bar{\Lambda}  \pi^+$ to clearly define the $\Sigma^-$ signal region;  therefore, one can finally reconstruct a $\Sigma^0$ in the rest of the event to represent  the $\Sigma^- \to \Sigma^0 e^- \bar{\nu}_e$ signal;   namely, the lepton pairs can be missed and reconstructed by examining the tagged signals.
  To study $\Xi^- \to \Xi^0  e^- \bar{\nu}_e$ in $J/\psi \to \Xi^-\bar{\Xi}^+$ decay, one can reconstruct the $\bar{\Xi}^+$ with $\bar{\Xi}^+ \to \bar{\Lambda}\pi^+$ decay as a "tag", then look at the $\Xi^-$ signal on the  recoiling mass of the fully reconstructed $\bar{\Xi}^+$,  and finally reconstruct a $\Xi^0 $ (with a $\Xi^0\to \Lambda \pi^0$ mode)   to represent the signals.   All these analyses will be benefit from the well-known center-of-mass energy of the initial $e^+e^-$ collision at BESIII/BEPCII.  The expected sensitivity on the branching fractions will be in the range of $10^{-5}$--$10^{-6}$.

   \begin{table}[htbp]
\centering
\caption{ Allowed baryon transitions $B_i \to B_f e \nu$ between members of the $J^P = \frac{1}{2}^+$ SU(3) baryon octet.
The present status of branching fractions for the semileptonic hyperon decays from PDG2016~\cite{pdg2016}  and data for the form-factor ratio $g_1(0)/f_1(0)$ in the Cabibbo model are also shown.  ``-'' indicates ``not available.'' }
 \label{tab:beta-decays}
\begin{tabular}{@{}llll}
\hline
    Decay mode    &   ${\cal B}$ ($\times 10^{-4}$)                 & $|\Delta S|$  &  $g_1(0)/f_1(0)$ \\
\hline
$\Lambda \to p e^- \bar{\nu}_e $ &    $8.32\pm0.14$    & 1  & $0.718\pm 0.015$\\
$\Sigma^+ \to \Lambda e^+ \nu_e$    & $0.20\pm0.05$   & 0 & - \\
$\Sigma^- \to n e^- \bar{\nu}_e$  & $10.17\pm 0.34$  & 1 & $-0.340 \pm 0.017$ \\
$\Sigma^- \to \Lambda e^- \bar{\nu}_e$  & $0.573\pm0.027$  & 0 & - \\
$\Sigma^- \to \Sigma^0 e^- \bar{\nu}_e$  & -  & 0 & - \\
$\Xi^0 \to \Sigma^+  e^- \bar{\nu}_e$  & $2.52\pm0.08$  & 1 &  $1.210 \pm 0.050$  \\
$\Xi^- \to \Lambda  e^- \bar{\nu}_e$  & $5.63\pm0.31$  & 1 & $0.250\pm 0.050$  \\
$\Xi^- \to \Sigma^0  e^- \bar{\nu}_e$  & $0.87\pm0.17$  & 1   & - \\
$\Xi^- \to \Xi^0  e^- \bar{\nu}_e$  & $<23$ (90\% C.L.)  & 0   & - \\
$\Omega^- \to \Xi^0 e^- \bar{\nu}_e $ & $56\pm28$  & 1 & - \\
\hline
\end{tabular}
\end{table}

   \begin{table}[htbp]
 \begin{center}
\caption{Present status of branching fractions (${\cal B}$) for the $\Delta S = -\Delta Q$ or $\Delta S =2$ rare semileptonic hyperon decays from PDG2016~\cite{pdg2016}. ``-'' indicates ``not available.'' }
 \label{tab:beta-decays-rare}
\begin{tabular}{@{}lll}
\hline
    Decay mode    &   ${\cal B}$ ($\times 10^{-6}$)                 & $\Delta S$ \\
           &   $@90\%$ C.L.                &  \\

\hline
$\Sigma^+ \to n e^+ \nu_e$$^*$    & $<5$   & 1 \\
$\Xi^0 \to \Sigma^-  e^+ \nu_e$$^*$  & $<900$  & 1   \\
$\Xi^0 \to p  e^- \bar{\nu}_e$  & $<1300$  & 2   \\
$\Xi^- \to n  e^- \bar{\nu}_e$  & $<3200 $  & 2   \\
$\Omega^- \to \Lambda e^- \bar{\nu}_e $ & -  & 2 \\
$\Omega^- \to \Sigma^0 e^- \bar{\nu}_e $ & -  & 2 \\
\hline
\end{tabular}
\vspace{-0.2cm}
  \begin{flushleft}
  \footnotesize{$^{*}$  $\Delta S = - \Delta Q$ process. }
   \end{flushleft}
    \end{center}
\end{table}

For the rare and forbidden semileptonic hyperon decays with $\Delta S = - \Delta Q$ or $\Delta S =2$,  examples are listed in Table~\ref{tab:beta-decays-rare},
with  available data from the PDG~\cite{pdg2016}, which were from experiments conducted nearly forty years ago.  Most of these upper limits can be improved with data from BESIII,  and the expected sensitivity ranges from $10^{-6}$ to $10^{-5}$ on these branching fraction measurements.

\section{Radiative hyperon decays}
\label{sec:radiative}

Since the discovery of hyperons, the nature of (weak) radiative decays  still remains an open question~\cite{Bernstein:1965hj,Lach:1995we}.
Study of weak radiative hyperon decays  provides an important tool for investigating the interplay of the  electromagnetic, weak, and strong interactions.
The description of these processes in terms of well-understood electroweak forces is complicated by the presence of strong interactions~\cite{Balitsky:1989ry,Gaillard:1985km}.
The hyperons in the baryon octet provide us with multiple reactions of this class with varying quark content of the initial and final state baryons. These decays are listed in Table~\ref{tab:radiative}.
\begin{table}[htbp]
 \begin{center}
\caption{Present status of branching fractions  (${\cal B}$) and asymmetry parameters ($\alpha_\gamma$) for the radiative hyperon decays from PDG2016~\cite{pdg2016}.
Neither the branching fraction  nor the asymmetry parameter
for $\Sigma^0 \to n \gamma$ has been measured owing to their huge electromagnetic partial widths.  ``-'' indicates ``not available.'' }
 \label{tab:radiative}
\begin{tabular}{@{}lll}
\hline
    $B_i \to B_f \gamma$    &   ${\cal B}$ ($\times 10^{-3}$)                  & $\alpha_{\gamma}$  \\
\hline
$\Lambda \to n \gamma$ &    $1.75\pm0.15$    & -\\
$\Sigma^+ \to p \gamma$  & $1.23\pm0.05$   & $-0.76\pm0.08$\\
$\Sigma^0 \to n \gamma$ & -  & - \\
$\Xi^0 \to \Lambda \gamma$ & $1.17\pm0.07$  & $-0.70\pm0.07$   \\
$\Xi^0 \to \Sigma^0 \gamma$ & $3.33\pm0.10$ &$-0.69\pm0.06$  \\
$\Xi^- \to \Sigma^- \gamma$ &   $0.127\pm0.023$ & $1.0\pm1.3$ \\
$\Omega^- \to \Xi^- \gamma $ & $<0.46$ (90\% C.L.)  & - \\
\hline
\end{tabular}
\end{center}
\end{table}

The transition matrix element $T$ for a general radiative decay of a hyperon $B_i$ of momentum $p$ to a baryon $B_f$ of momentum $p^\prime$ and a photon momentum $q$,
\begin{eqnarray}
B_i(p) \to B_f(p^\prime) + \gamma(q),
\label{cross:gen-rad}
\end{eqnarray}
is given by~\cite{Zenczykowski:2005cs}
\begin{eqnarray}
T = G_F\frac{e}{\sqrt{4\pi}}\epsilon_\nu \bar{u}(p^\prime)({\bf A}+{\bf B}\gamma_5) \sigma_{\mu\nu}q_\mu u(p),
\label{eq:t-rad}
\end{eqnarray}
where $\bar{u}(p^\prime)$ and $u(p)$ are the spinor wave functions of the baryon and hyperon, respectively, $\epsilon_\nu$ is the polarization vector of the photon, ${\bf A}$ and ${\bf B}$ are the parity-conserving (M1) and parity-violating (E1) amplitudes, $  \sigma_{\mu\nu}$ and $\gamma_5$ are the combinations of the
Dirac gamma matrices, $G_F$ is the Fermi constant, and $e$ is the electron charge.
The asymmetry parameter  is~\cite{Borasoy:1999nt}
\begin{eqnarray}
\alpha_\gamma =  \frac{2{\rm Re}({\bf A}^*{\bf B})}{|{\bf A}|^2+|{\bf B}|^2}.
\label{eq:t-rad-a}
\end{eqnarray}
One needs both nonzero ${\bf A}$ and ${\bf B}$ amplitudes to get nonzero asymmetry.

For a polarized spin-1/2 hyperon decaying radiatively via a $\Delta Q=0$, $\Delta S=1$ transition, which is a flavor-changing-neutral-current (FCNC) process, the angular distribution of the direction $\hat{\bf p}$ of
the final spin-1/2 baryon in the hyperon rest frame is
\begin{eqnarray}
\frac{dN}{d\Omega} = \frac{N^0}{4\pi}(1+\alpha_\gamma {\bf P}_i \cdot \hat{\bf p}),
\label{eq:t-rad-g}
\end{eqnarray}
where ${\bf P}_i$ is the polarization of the decaying hyperon. If the decaying hyperon is unpolarized, the decay baryon has a longitudinal polarization given by
$P_f= - \alpha_\gamma$~\cite{Behrends:1958zz}.

There is an old theorem by Hara~\cite{Hara:1964}  regarding the vanishing of the parity-violating ${\bf B}$ amplitudes for  weak radiative hyperon decays.  According to Hara's theorem,
 the relevant parity-violating amplitude should vanish in the SU(3) limit. For broken SU(3), according to the size of hadron-level SU(3)-breaking effects elsewhere, one would expect this asymmetry to be of the order  of $\pm0.2$~\cite{Zenczykowski:2005cs}, and not of the order of $-1$ as indicated in Table~\ref{tab:radiative}. The situation was further confounded by a number of theoretical calculations that violated Hara's theorem (even) in the SU(3) limit~\cite{Lach:1995we}.   In particular, the large asymmetry observed in the decay $\Sigma^+ \to p \gamma$  has been fueling numerous discussions over many years and remains poorly understood at the parton level~\cite{Lach:1995we}.

  As listed in Table~\ref{tab:radiative},  the uncertainty on the asymmetry parameter in the $\Xi^- \to \Sigma^- \gamma$  decay is still very large and can be improved, and there is no measurement of the decay parameter for the $\Lambda \to n \gamma$ decay yet. The current data were all from the fixed target experiments more than a few decades ago,  and improved or cross-checked measurements should be done in the future.  BESIII will collect $\sim$10 billion $J/\psi$ events, and the numbers of hyperon events from the $J/\psi$ decays are estimated in Tables~\ref{tab:2} and~\ref{tab:3}.

    In addition to the two-body radiative decays, weak hyperon dilepton decays will provide additional information on the radiative transitions~\cite{Bos:1996ig, Martemyanov:2003ca} .  According to PDG2016, the only observed hyperon dilepton decay is the $\Xi^0 \to \Lambda e^+e^-$ decay, which is measured to be ${\cal B}(\Xi^0 \to \Lambda e^+e^-) = (7.6\pm 0.6)\times 10^{-6}$~\cite{Batley:2007hp},   which is consistent with an inner bremsstrahlung-like production mechanism for the $e^+e^-$  pair. The consistency is further supported by the $e^+e^-$ invariance mass spectrum. The decay parameter was determined to be $\alpha_{\Xi\Lambda ee} = -0.8 \pm 0.2$~\cite{Batley:2007hp}, which is consistent with that measured for the $\Xi^0 \to \Lambda \gamma$ decay as listed in Table~\ref{tab:radiative}.  A detailed discussion of radiative dilepton decays is presented in the next section.

\section{Rare and forbidden  hyperon decays}
\label{sec:rare}

\subsection{$B_i \to B_f l^+l^-$ dilepton decays}

In the Type A region as listed in Table~\ref{tab:rare-f},
the decays $B_i \to B_f l^+l^-$ (where $l = e$, $\mu$; i.e., Dalitz decay) can be described as proceeding through both short-distance and long-distance contributions. In the SM, the leading short-distance contribution comes from an FCNC interaction,   which is allowed only at loop level~\cite{He:2005yn}.   As discussed in Section~\ref{sec:radiative}, the decays of $B_i \to B_f e^+e ^-$  play an important role in helping us to understand the dynamics of radiative hyperon decays, and  the decay rates  are suppressed by two orders of magnitudes if we assume an inner bremsstrahlung-like mechanism producing the $e^+e^-$ pairs~\cite{Bergstrom:1987wr}.  For the  decay $B_i \to B_f  \mu^+\mu^-$, the process is dominated by long-distance contributions,  which are from the $B_i \to B_f  \gamma^*$ and $B_i \to B_f  V^*$ (where $V$ could be $\rho$/$\omega$ vector mesons) processes, namely, the so-called vector dominant model mechanism~\cite{Zenczykowski:2005cs}.  For example,  the branching fraction of $\Sigma^+ \to p \mu^+\mu^-$ is predicted to be ${\cal B}(\Sigma^+ \to p \mu^+\mu^-) \in [1.6, 9.0]\times 10^{-8}$~\cite{He:2005yn} by considering the long-distance contributions, while the short-distance SM contributions are suppressed at a branching fraction of $\sim$$10^{-12}$.
Dalitz decays of hyperons are of particular interest, as they also allow a direct search for a new scalar or vector particle, which could lead to an $s \to d$ transition at the tree level~\cite{Gorbunov:2000cz}.  Recently, the Large Hadron Collider beauty experiment (LHCb) searched for $\Sigma^+ \to p \mu^+\mu^-$ decay and observed an excess of events with respect to the background expectations with a signal significance of 4.0 standard deviations. No significant structure is observed in the dimuon invariant mass distribution.
Owing to the difficulty of normalization in the absence of signal,  only an upper limit on the branching fraction is set to be $6.3\times 10^{-8}$ at 95\% C.L.~\cite{Dettori:2016vku}, which agrees with the SM prediction~\cite{He:2005yn},
whereas the sensitivities for the decays $B_i \to B_f l^+l^-$ are  estimated to be $10^{-6}$--$10^{-7}$ as listed in Table~\ref{tab:rare-f} (Type A) with the BESIII data.

\begin{table}[htbp]
  \begin{center}
\caption{Rare and forbidden hyperon decays and expected sensitivities with  $10^{10}$ events on the  $J/\psi$
peak and  $3\times 10^{9}$ events on the $\psi(2S)$ peak.  Type A decay modes are through a photon--penguin-like weak neutral current;
Type B decay modes are through a $Z$--penguin-like weak neutral current; Type C decay modes are  neutrinoless double $\beta$ decays, which violate lepton number. The current data are from the world average in PDG2016~\cite{pdg2016}. ``-'' indicates ``not available.''  }
 \label{tab:rare-f}
\begin{tabular}{@{}llll}
\hline
   Decay mode    & Current data &   Sensitivity             \\
        & ${\cal B}$($\times 10^{-6}$) &  ${\cal B}$(90\%C.L.)  ($\times 10^{-6}$)  &     Type        \\
\hline
$\Lambda \to n e^+ e^-$  &    -    & $<0.8$ \\
$\Sigma^+ \to p e^+e^- $  & $<7$   & $<0.4$\\
$\Xi^0 \to \Lambda e^+e^-$ & $7.6\pm0.6$  & $<1.2$   \\
$\Xi^0 \to \Sigma^0 e^+e^-$ & - &$<1.3$  \\
$\Xi^- \to \Sigma^- e^+e^-$ &   - & $<1.0$ \\
$\Omega^- \to \Xi^- e^+e^- $ & - & $<26.0$ & Type A \\

$\Sigma^+ \to p \mu^+\mu^- $  & $(0.09^{+0.09}_{-0.08})$   & $<0.4$\\
$\Omega^- \to \Xi^-  \mu^+\mu^- $ & - & $<30.0$\\

 \hline
$\Lambda \to n \nu\bar{\nu} $  &    -   & $<0.3$\\
$\Sigma^+ \to p \nu\bar{\nu} $  & -   & $ <0.4$\\
$\Xi^0 \to \Lambda \nu\bar{\nu}$ & - & $<0.8$   \\
$\Xi^0 \to \Sigma^0 \nu\bar{\nu}$ & - &$<0.9$  & Type B \\
$\Xi^- \to \Sigma^- \nu\bar{\nu}$ &  - & -$^*$ \\
$\Omega^- \to \Xi^- \nu\bar{\nu} $ & -  & $<26.0$ \\
\hline
$\Sigma^- \to \Sigma^+ e^- e^-  $  & -  & $<1.0$\\
$\Sigma^- \to p e^- e^-  $  & -  & $<0.6$\\
$\Xi^- \to p e^-e^- $ &   - & $<0.4$ \\
$\Xi^- \to \Sigma^+ e^-e^- $ &   - & $<0.7$ \\
$\Omega^- \to \Sigma^+ e^-e^- $  & -  & $<15.0$  & \\
$\Sigma^- \to p \mu^- \mu^-  $  & -  & $<1.1$\\
$\Xi^- \to p \mu^-\mu^- $ &   $<0.04$ & $<0.5$  & Type C \\
$\Omega^- \to \Sigma^+ \mu^-\mu^- $  & -  & $< 17.0$ & \\
$\Sigma^- \to p e^- \mu^-  $  & -  & $<0.8$\\
$\Xi^- \to p e^-\mu^- $ &   - & $<0.5$ \\
$\Xi^- \to \Sigma^+ e^-\mu^- $ &  - & $<0.8$ \\
$\Omega^- \to \Sigma^+ e^-\mu^- $ &-  & $<17.0$ \\

\hline
\end{tabular}
\vspace{-0.2cm}
  \begin{flushleft}
  \footnotesize{$^{*}$  It is hard to reconstruct the $\Xi^- \to \Sigma^- \nu\bar{\nu}$ decay since the $\Sigma^-$ decays into final states including neutron. Thus both neutron and neutrinos can not be detected at BESIII, and the "tag technique" will not work. }
   \end{flushleft}

\end{center}
\end{table}

\subsection{$B_i \to B_f \nu\bar{\nu}$ decays via a $Z$-type penguin }

Analogous to the rare  $K\to \pi \nu \bar{\nu}$ decays~\cite{Marciano:1996wy}, the rare $ B_i \to B_f \nu\bar{\nu}$ decays are important tools for testing the SM and for searching for possible new physics. As they proceed through an FCNC process, though they are very suppressed in the SM, they show an exceptional sensitivity to short-distance physics that is similar  to the $K\to \pi \nu \bar{\nu}$ decays~\cite{Buras:2005gr,Buras:2004uu}.    There are many discussions on the new physics models that may
contribute to the $s \to d \nu \bar{\nu}$ transition at the loop or even tree level; for a recent review, see Ref.~\cite{Buras:2015yca}.  Unfortunately, no $B_i \to B_f \nu\bar{\nu}$ decays  have been searched for experimentally so far.   As an example, assuming that the decay $\Sigma^+ \to  p \nu \bar{\nu}$ is dominated by the short-distance contribution,  one can make an estimation based on the following relation~\cite{feng-xu2016,Marciano:1996wy}:
 \begin{eqnarray}
\frac{\Gamma(K^+ \to \pi^+ \nu \bar{\nu})}{\Gamma(K^+ \to \pi^0 e^+ \nu)} \approx  \frac{\Gamma(\Sigma^+ \to p \nu \bar{\nu})}{\Gamma(\Sigma^- \to n e^+ \nu)}.
\label{eq:short-estim}
\end{eqnarray}
Thus,  in the SM, one can estimate  the branching fraction to be ${\cal B} (\Sigma^+ \to p \nu \bar{\nu}) \sim 5\times 10^{-13}$ by considering the lifetime difference between $\Sigma^+$ and $\Sigma^-$ and  isospin rotation~\cite{feng-xu2016}.    Obviously, BESIII cannot reach a sensitivity of $10^{-13}$.

Since the neutrinos in the final states  are ``invisible'' in the detector  and, at BESIII, the initial energy and momentum of electron and positron are known,  one can use the ``tag technique'' to fully reconstruct one of the hyperons and look at the recoil side in the $\psi$ decay into hyperon pairs.  It would be  interesting to make a first study of the $\Sigma^+ \to p \nu \bar{\nu}$ decay, and one should obtain a sensitivity of $10^{-8}$ with $10^{10}$ $J/\psi$ decay events.   More examples of these decays are listed in Table~\ref{tab:rare-f} (Type B), which are unique accomplishments of the BESIII experiment.

\subsection{Lepton-number-violating  decays with $\Delta L =2$}
\label{subsec:lnv}

Nonzero neutrino mass is now well established~\cite{neutrinos-1,neutrinos-2,neutrinos-3,An:2012eh}; therefore, it is crucially important to study the properties of these nonzero-mass neutrinos. One interesting question is whether neutrinos are Majorana or Dirac neutrinos~\cite{Pontecorvo:1957qd, Gribov:1968kq}.
 Lepton-number-violating (LNV) interactions with $\Delta L =2$ are widely viewed as the most robust test of the Majorana nature of massive neutrinos~\cite{Pontecorvo:1957qd, Gribov:1968kq}.  Currently,  neutrinoless double $\beta$ ($0\nu\beta\beta$) decays of  heavy nuclei~\cite{Rodejohann:2011mu} have become the most sensitive way to search for the effects of very light Majorana neutrinos. However, theoretical analysis of these processes is complicated  by the  nuclear matrix elements, which are very difficult to calculate reliably.   Here we focus on the $\Delta L = 2$ transitions between spin-1/2 hyperons,
 $B^-_i \to B^+_f l^- l^{\prime -}$ (where $l$, $l^\prime = e$ or $\mu$).  Examples of these decays are listed in Table~\ref{tab:rare-f} (Type C).

 Only one experimental upper limit for the channels listed in Table~\ref{tab:rare-f} (Type C) has been reported so far, namely,
 ${\cal B}(\Xi^- \to p \mu^-\mu^-) < 4.0 \times 10^{-8}$ .  In the case of the decays listed in Table~\ref{tab:rare-f} (Type C),
  two down-type (d or s) quarks convert into two up-quarks, changing the charge of hyperons according to the $\Delta Q = \Delta L = +2$ rule.
  These quark transitions  are assumed to occur at the same space-time location and, therefore, they are driven by local four-quark operators~\cite{Barbero:2002wm,Barbero:2007zm,Barbero:2013fc}.  Thus the study of the relatively simpler case provided by $0\nu\beta\beta$ hyperons decays may shed some light on the approximations used to evaluate the hadronic matrix elements relevant for similar nuclear decays~\cite{Barbero:2002wm,Barbero:2007zm,Barbero:2013fc}.
  In Ref.~\cite{Barbero:2007zm}, based on the model where $\Sigma^- \to p e^-e^-$ decay is induced by a loop of baryons and light Majorana neutrinos,
  the predicted branching fraction for $\Sigma^- \to p e^-e^-$ decay, as an example, is $<$$10^{-33}$,  whereas, in Ref.~\cite{Barbero:2013fc},  based on a model
  with four-quark operators,  the predicted branching fraction is $<$$10^{-23}$, which is still too small to be observed in any future high-intensity experiment.
  Thus, any observable  rates for these decays must indicate a new interaction beyond the SM.

  It will be very interesting to search for the  $B^-_i \to B^+_f e^-e^-$ decays at BESIII with $10^{10}$ $J/\psi$ decay into hyperon--antihyperon final states, which will produce $10^{7}$--$10^{8}$ hyperons.  Therefore, the sensitivities will be on the order of $10^{-7}$ with clean backgrounds.  Note that the decays of $B^-_i \to B^+_f \mu^-\mu^-$ can be also searched for at LHCb with higher sensitivities owing to the huge production cross section there.

 \begin{table}[htbp]
  \begin{center}
\caption{ Lepton- or baryon-number-violating  hyperon decays and expected sensitivities with  $10^{10}$ events on the  $J/\psi$
peak and  $3\times 10^{9}$ events on the $\psi(2S)$ peak.  The current data are from CLAS~\cite{McCracken:2015coa} as listed in PDG2016~\cite{pdg2016}. ``-'' indicates ``not available,''  $l = e$ or $\mu$, and $M^{\pm}$ refers to the charged stable mesons ($M^\pm= \pi^\pm$ or $K^\pm$).
Each reaction shows evidence of $\Delta L = \pm 1$ or/and $\Delta B \neq 0$, and each reaction conserves electric charge and angular momentum.  }
 \label{tab:rare-exotics}
\begin{tabular}{@{}lllll}
\hline
   Decay mode    & Current data &   Sensitivity       &$\Delta L$  & $\Delta B$       \\
        & ${\cal B}$ ($\times 10^{-6}$) & ${\cal B}$   ($\times 10^{-6}$)        \\
              & & (90\%C.L.)         \\
\hline
$\Lambda \to M^+ l^-$  &   $<0.4\sim 3.0$~\cite{McCracken:2015coa}   & $<0.1$   & +1 & $-1$ \\
$\Lambda \to M^- l^+$  &    $<0.4 \sim 3.0$~\cite{McCracken:2015coa}    &  $<0.1$ & $-1$ & $-1$ \\
$\Lambda \to K_S  \nu $  &    $<20$~\cite{McCracken:2015coa}   & $<0.6$   & $+1$ & $-1$ \\
$\Sigma^+ \to K_S  l^+ $       & -   & $<0.2$  & $-1$ & $-1$ \\
$\Sigma^- \to K_S  l^- $       & -   &  $<1.0$ & $+1$ & $-1$ \\
$\Xi^- \to K_S  l^- $       & -   &$<0.2$   & $+1$ & $-1$ \\
$\Xi^0 \to M^+  l^-$ & -  &  $<0.1$& +1 & $-1$ \\
$\Xi^0 \to M^- l^+$  &  -   & $<0.1$  & $-1$ & $-1$ \\
$\Xi^0 \to K_S  \nu $  &    -   & $<2.0$  & $+1$ & $-1$ \\

\hline
\end{tabular}
\end{center}
\end{table}

\subsection{Other decays violating lepton number and  baryon number }

The SM has been proved to be extremely successful. As we discussed in Subsection~\ref{subsec:lnv},  the nonzero neutrino mass indicates that the lepton number ($L$) may be violated.  There are also a number of reasons to consider baryon number violation: 1. there is a suggestion originally from Sakharov theory that $CP$ violation
combined with a  baryon-number-violating (BNV)  interaction can explain the baryon asymmetry of the universe~\cite{Sakharov:1967dj}; 2. many grand unified theories allow the proton to decay~\cite{Georgi}; 3. $B$--$L$ is an important symmetry and, therefore, if $\Delta L$ exists, there is also a $\Delta B$ interaction. In particular, in the SM, in the electroweak phase transition, there is a possibility of a $\Delta B$, $\Delta L$ transition that conserves $B$--$L$, an essential component of theories of leptogenesis ~\cite{An:2009vq}.  Recently, searches for the BNV and LNV decays of $\Lambda$ were performed by the CLAS experiment~\cite{McCracken:2015coa} by using a data set for photoproduction off of the proton collected with the CLAS detector at Jefferson Laboratory containing $\sim$$1.8\times 10^6$  reconstructed $\gamma p \to K^+ \Lambda$  events. The sensitivities on the branching fraction for each of the processes studied ranged from
$7\times 10^{-7}$ to $2\times 10^{-5}$~\cite{McCracken:2015coa}.  All BNV and LNV decays can be further studied with data from $J/\psi$ decay into hyperon pairs as listed in Table~\ref{tab:rare-exotics}.  Some of them ($\Sigma$ and $\Xi$ hyperon decays) were never searched for, and the sensitivities will be $10^{-7}$ or  better.   In addition, baryon-number violation with $\Delta B =2$ processes can be searched for in the $\Lambda \to \bar{p} \pi^+$ decays that were also presented by the CLAS experiment~\cite{McCracken:2015coa} for the first time.  At BESIII, possible baryon-number-violation  can be probed in $J/\psi \to \Lambda \bar{\Lambda}$ decay with better sensitivity than that of CLAS. Furthermore,  $\Lambda$--$\bar{\Lambda}$ oscillations can be investigated by using coherent productions of $\Lambda$ pairs in   $J/\psi \to \Lambda \bar{\Lambda}$ decay; for details, see Ref.~\cite{Kang:2009xt,Berezhiani:2015uya}.

In conclusion, the search for baryon and lepton number nonconservation
represents an important probe of physics beyond the SM and particularly new physics at a very high mass scale.

\section{Summary}
\label{sec:sum}

Ever since the discovery of hyperons and of their weak decays it has been a challenge to measure with precision the properties of their decays and to test the SM and beyond.
Hyperon semileptonic  and radiative decay are still a challenge on both experimental and theoretical fronts.   The rare and forbidden hyperon decays will play important role in the search for new physics~\cite{pdg2016,Chao:1990hg, He:1995na,Donoghue:1986hh,Chang:1994wk,He:1994vu,Tandean:2003fr,Tandean:2002vy,Tandean:2004mv,He:1992ng,Kang:2010td,Abdesselam:2016nym}.
In this paper, we propose studying hyperon decays at BESIII with events in the J$/\psi$ or $\psi(2S)$  decay into hyperon pairs.
We learned that the two-body decays of the $J/\psi$ and $\psi(2S)$ will provide a pristine experimental environment;  especially, rare decays and decays with invisible final states can be probed by using  the ``tag technique'' with the $e^+ e^-$ collision experiment.  With one year's integrated luminosity at BESIII, $\sim$$10^{6}$--$10^8$ hyperons, $\Lambda$, $\Sigma$, $\Xi$, and $\Omega$, will be produced in the J$/\psi$ and $\psi(2S)$ decays.  Based on those samples,  the sensitivity for the  measurements of the branching fractions of the hyperon decays is in the range of $10^{-5}$--$10^{-8}$. Study of hyperon decays will no doubt prove be an unexpected and rewarding research field at BESIII.

    In addition, the hyperon program will be one of the strong motivations for collecting data at the $J/\psi$ and $\psi(2S)$ peaks at the planned super-$\tau$-charm factory~\cite{Bondar:2013cja,Zhou:2016qfu}.
  Meanwhile, the decay modes with $\mu^\pm$ final states can be probed using the LHCb, where the production cross sections are huge~\cite{Dettori:2016vku}.

     {\it Discussion:}  We need theoretical input for these rare decays; for example,  the radiative dilepton decays $B_i \to B_f l^+ l^-$ should be revisited with
the recent development of  chiral perturbation theory~\cite{Kimura:2016xnx}.  More theoretical efforts should be focused on the $\Delta L =2$ LNV decays  $B^-_i \to B^+_f l^- l^{\prime -}$, which may have an impact on massive neutrinos~\cite{Dong:2013raa}.   There are no theoretical estimations of the $B_i \to B_f \nu\bar{\nu}$ decays, and efforts on the analogous $K^+ (K_L) \to \pi^+ (\pi^0)  \nu \bar{\nu}$ decay could be applied to hyperon decays, for example, to the recent lattice QCD calculations on the $Z$--penguin-like  kaon decays~\cite{Christ:2016eae,Christ:2016psm,Christ:2015aha}.

     Note that the estimations of the sensitivities at BESIII are all based on educated guesses of the detection efficiencies and thus may deviate from the
     true values; more studies with correct angular distributions~\cite{Lee:1957qs,Kadeer:2005aq} should be done in the future.

\section*{Acknowledgments}
The author would like to thank Stephen L. Olsen, I.~I.~Bigi, and Xu Feng for useful  discussions and suggestions and also  J.~G.~K\"orner, Francesco Dettori, and J. Tandean for their useful comments.   This
work is supported in part by the National Natural Science
Foundation of China under Contracts Nos.
11335009 and 11125525, the Joint Large-Scale Scientific Facility Funds of the NSFC and CAS under Contract No.  U1532257, CAS under Contract No. QYZDJ-SSW-SLH003, and
the National Key Basic Research Program of China under Contract No. 2015CB856700.




\end{document}